\documentclass[lettersize,journal]{IEEEtran}
\usepackage{amsmath,amsfonts}
\usepackage{algorithmic}
\usepackage{algorithm}
\usepackage{array}
\usepackage[caption=false,font=normalsize,labelfont=sf,textfont=sf]{subfig}
\usepackage{textcomp}
\usepackage{stfloats}
\usepackage{url}
\usepackage{balance}
\usepackage{verbatim}
\usepackage{graphicx}
\usepackage{cite}
\usepackage{amssymb}
\usepackage[dvipsnames]{xcolor}
\begin{document}

\title{DFPF-Net:Dynamically Focused Progressive Fusion Network for Remote Sensing Change Detection}

\author{Chengming~Wang, Peng Duan, Jinjiang Li
\thanks{C. Wang, P. Duan and J. Li are with School of Computer Science and Technology, Shandong Technology and Business University, Yantai 264005,
			China}
	
}

\markboth{Journal of \LaTeX\ Class Files,~Vol.~14, No.~8, August~2021}%
{Shell \MakeLowercase{\textit{et al.}}: A Sample Article Using IEEEtran.cls for IEEE Journals}

\IEEEpubid{0000--0000/00\$00.00~\copyright~2021 IEEE}

\maketitle

\begin{abstract}
Change detection (CD) has extensive applications and is a crucial method for identifying and localizing target changes. In recent years, various CD methods represented by convolutional neural network (CNN) and transformer have achieved significant success in effectively detecting difference areas in bi-temporal remote sensing images. However, CNN still exhibit limitations in local feature extraction when confronted with pseudo changes caused by different object types across global scales. Although transformers can effectively detect true change regions due to their long-range dependencies, the shadows cast by buildings under varying lighting conditions can introduce localized noise in these areas. To address these challenges, we propose the dynamically focused progressive fusion network (DFPF-Net) to simultaneously tackle global and local noise influences. On one hand, we utilize a pyramid vision transformer (PVT) as a weight-shared siamese network to implement change detection, efficiently fusing multi-level features extracted from the pyramid structure through a residual based progressive enhanced fusion module (PEFM). On the other hand, we propose the dynamic change focus module (DCFM), which employs attention mechanisms and edge detection algorithms to mitigate noise interference across varying ranges. Extensive experiments on four datasets demonstrate that DFPF-Net outperforms mainstream CD methods.
\end{abstract}


\section{INTRODUCTION}
\IEEEPARstart{C}{hange}  detection (CD) in remote sensing images is the process of detecting changes in surface objects within the same geographical area using bi-temporal remote sensing images. In CD techniques, binary change detection (BCD) distinguishes change area pixels using binary classification images, in this paper, we focus on the BCD approach. Remote sensing (RS) has extensive surface coverage and strong observational capabilities. In recent years, as satellite resolution improves, RS technology has been successfully integrated with various CD methods and has been applied in fields such as urban planning, land use surveys, disaster mapping, and forest cover mapping \cite{1,2,3,4}. This has significant implications for urban development, disaster prevention, and information statistics.

In traditional change detection methods, pixel-level change detection treats independent pixels as detection units, analyzing differences in bi-temporal remote sensing images through arithmetic operations, including methods such as image differencing, regression analysis, and independent component analysis (ICA) \cite{53,54,55}. Feature-level CD achieves change detection by comparing features such as texture, edges, and spatial structures of bi-temporal RS images. EDRCNN integrates discriminative information and a priori knowledge of edge structure into a single framework for edge guidance \cite{6}. Gong et al. \cite{5} proposed a method utilizing two difference images constructed from intensity and texture information. In this approach, robust principal component analysis (RPCA) is employed to extract texture differences by separating irrelevant and noisy elements from Gabor responses, effectively enhancing the representation of texture features. Scene-level change detection understands semantic information by analyzing different attributes of multiple objects in surface information and their spatial distribution. SCDTN employs pre-training combined with change vector analysis for scene-level pre-detection, followed by pixel-level classification using decision trees \cite{7}. Simple and rapid traditional methods \cite{8} can only extract shallow feature information from images when facing complex features, lacking understanding of deep information and change comparisons. Machine learning methods are more flexible and efficient compared to earlier traditional methods, enabling automatic learning and feature extraction from data, thereby effectively processing complex data information. Support vector machines (SVM) separate change areas from non-change areas by constructing an optimal hyperplane and use kernel functions to handle nonlinear data \cite{9,10}. Decision trees classify data based on features by constructing tree models. Random forests enhance classification accuracy through a voting mechanism by integrating decision tree ideas, thus handling complex features \cite{11,12}. K-means clustering divides pixels into k clusters to identify areas of change difference \cite{13}. Although these machine learning methods can gradually understand deep information in images to detect change differences, they still lack sufficient automatic learning capability to accurately detect pseudo-changes in surface images over different periods influenced by natural weather and do not effectively address noise such as shadows surrounding similar objects.

\IEEEpubidadjcol

The rapid development of deep learning has further enhanced the advantages of machine learning in CD, this has led to the development of various deep learning-based methods. The end-to-end learning approach simplifies the traditional CD process, reducing the need for manual intervention and feature engineering. Convolutional neural networks (CNN) possess strong local feature extraction capabilities, effectively capturing details such as edges, textures, and shapes in images. This makes CNN excel in handling richly detailed RS images, as their deep structures support the layer-wise extraction of features from low-level to high-level, enhancing their ability to detect difference areas in complex images. U-Net \cite{14} improves the accuracy of CD by combining features at different scales, and its powerful scalability has inspired the design and application of more methods based on U-shaped architectures in the field of CD \cite{15,16,17}. The self-attention mechanism of transformer \cite{18} demonstrates global capabilities in processing time series, achieving promising results. The widespread application of vision transformer (ViT) \cite{19} in computer vision has further encouraged their application in image processing tasks. The self-attention mechanism of ViT focuses on the dependencies between different positions in images, capturing global feature information of bi-temporal RS images to dynamically identify the changes between images. Pyramid ViT \cite{20} combines a pyramid structure with the global dynamic attention mechanism of ViT, endowing the model with multi-level feature extraction capabilities. PVTv2 \cite{21} introduces linear space-reducing attention to further lower computational costs and enable effective modeling, however, it performs poorly in handling detailed changes and regions with strong noise interference.

\begin{figure}[!t]
	\centering
	\includegraphics[width=3.5in]{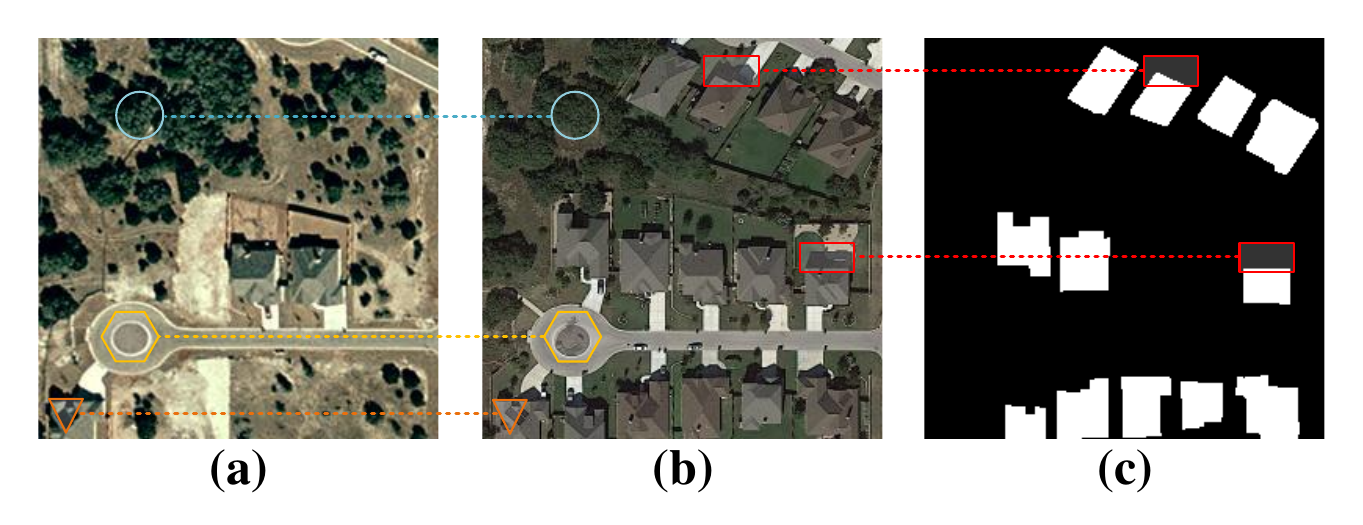}
	\caption{Illustrates an overview of the noise effects in CD. Subfigures (a), (b), and (c) show the bi-temporal RS images and the ground truth images after CD. Specifically, the comparison between (a) and (b) highlights the pseudo-change noise from different types of objects, such as trees and buildings in unchanged areas, while the comparison between (b) and (c) emphasizes the shadow noise from buildings in the areas of truly change.
	}
	\label{fig1}
\end{figure}

In RSCD tasks, pseudo-change interference is an inherent challenge, particularly due to the diversity of pseudo-changes. We consider large-scale factors such as weather, seasonal variations, and lighting as global noise, while building shadows are regarded as local noise. Our goal is to mitigate the impact of global noise and focus on suppressing local noise characterized by building shadows. Thus, we designed the dynamically focused progressive fusion network (DFPF-Net). First, we employed PVT as the backbone network to extract multi-level features from bi-temporal RS images. Additionally, we proposed PEFM to integrate image information, reducing the impact of pseudo-changes and building shadow noise through progressive fusion. Preprocessing is performed on the two images to accomplish preliminary denoising before the progressive fusion, followed by interpolation and absolute value operations on the multi-level features extracted from the pyramid structure to generate a weight image representing change intensity. This approach allows for clear identification of changes in the same area of the two images at different time points. A larger absolute difference value indicates a more pronounced change area, aiding subsequent feature extraction and fusion to enhance CD performance. In the subsequent fusion phase, we applied the concept of cross-attention to process the images at different levels. In PEFM, we employed a dual residual structure within the overall architecture to ensure the stability and coherence of the progressive fusion. We designed the DCFM to enhance focus on pseudo-change areas and building edge areas affected by shadows in bi-temporal RS images. For pseudo-change areas, we utilized attention mechanisms to emphasize the differences in these regions. Different weather conditions can lead to varying shadow characteristics around buildings, we applied edge detection methods to mitigate the impact of shadow regions on the detection of differences in target areas. The efficient global information modeling capability of agent attention \cite{22}, along with its low computational overhead, is combined with edge detection methods \cite{23} to reallocate image weights through residual structures and nonlinear activation functions, achieving a key step in detecting differences in bi-temporal RS images.

The main contributions of this work are summarized as follows:

\begin{itemize}
\item We developed a novel DFPF-Net to enhance the performance of CD on bi-temporal RS images. Our method is more effective compared to mainstream CD methods.

\item We proposed an innovative PEFM, which utilizes a progressive fusion approach with a residual structure to phase-wise process shallow and deep feature information, establishing strong associations to address diverse change scenarios. To effectively suppress noise, we designed the DCFM to clearly distinguish pseudo-change areas while mitigating the impact of shadows around buildings.

\item Extensive experiments on four public datasets show that DFPF-Net outperforms mainstream methods, delivering strong performance with reduced computational costs.

\end{itemize}

\begin{figure*}[!t]
	\centering
	\includegraphics[width=7in]{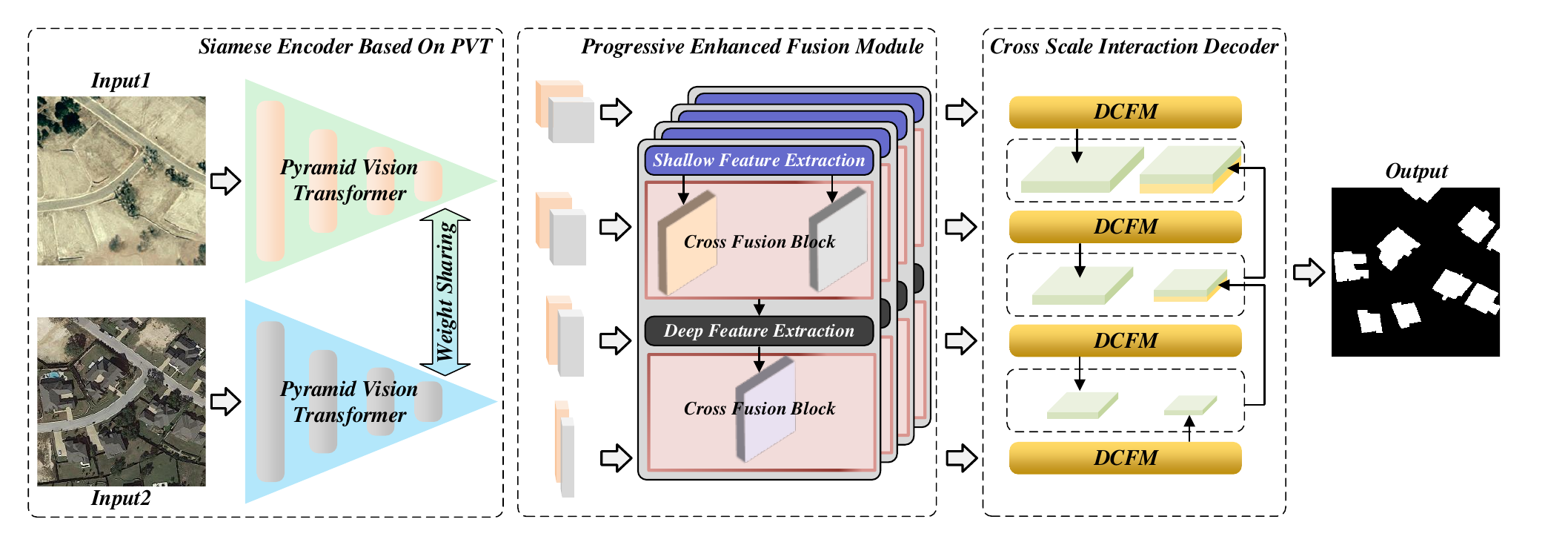}
	\caption{Provides an overview of the DFPF-Net. To process bi-temporal images, we first employ PVT to construct a weight-sharing siamese structure as the encoder. Then, we input the multi-scale feature information into the PEFM. Subsequently, we propose the DCFM to enhance the weights of the change regions, and finally, we perform differential information fusion using a cross scale interaction decoder.}
	\label{fig2}
\end{figure*}

\section{RELATED WORK}
\subsection{Traditional Methods in Change Detection}
With advancements in sensors and breakthroughs in information technology, high-resolution RS images obtained through cutting-edge technology have garnered widespread attention in change detection. BCD is commonly used methods in RSCD, implemented from the perspectives of pixel difference calculation. To detect changes in bi-temporal RS images, pixel-level change detection is achieved using differential image processing methods, which involve setting predefined thresholds based on the pixel values of the resulting differential images \cite{24}. Morphological feature extraction and analysis are employed to match points of interest in multi-temporal buildings, reducing the sensitivity issues associated with changes detected under varying imaging conditions \cite{25,26}. Conventional machine learning algorithms, such as support vector machines (SVM), K-Means clustering, and random forests, are applied in RSCD from various perspectives, leveraging their respective algorithmic advantages to improve the efficiency and accuracy of CD technologies in the RS field. However, traditional CD algorithms rely on manually designed spatiotemporal features, which can be resource-intensive when dealing with large-scale data collections, and these methods typically exhibit weaker robustness.

\subsection{CD Methods Based on CNN}
The advancements in deep learning technologies have led to a flourishing development of methods based on CNN, resulting in numerous RSCD techniques. The simplicity and flexibility of the U-Net architecture not only achieve significant advantages in image segmentation but also inspire various CD methods. Notably, FC-EF \cite{27} proposes three fully convolutional neural networks that input the concatenated bi-temporal images into U-Net for feature extraction, followed by three different processing approaches to detect feature map differences. STANet \cite{28} introduces a multi-scale self attention module that simulates spatiotemporal relationships by computing attention weights across different times and locations, generating discriminative features that capture spatiotemporal dependencies. IFNet \cite{29} employs a deep supervision method, feeding the extracted deep features into a differential discriminative network to enhance the boundary integrity and internal compactness of the detected targets. SNUNet \cite{30} emphasizes the importance of shallow information containing fine-grained features, proposing the use of densely connected siamese networks for compact information transfer to alleviate the loss of spatial information in deeper layers of the neural network.

In RSCD, the variability of change regions due to seasonal variations and complex geographical information inevitably necessitates deep networks for extracting high-level features. However, traditional deep networks often face issues such as gradient vanishing and exploding. The residual structure proposed by ResNet \cite{31} effectively addresses this problem, while MRA-SNet \cite{32} employs a multi-scale residual structure to extract spatial and spectral features at different scales, capturing detailed feature information at deeper levels.

Despite the promising performance of these CNN-based methods, which provide innovative approaches for processing and analyzing RS images and advance the application of CD technologies in this field, the limitations of CNN in local feature extraction hinder their ability to effectively fuse bi-temporal RS image information. Moreover, they fall short of leveraging the advantages of attention mechanisms in acquiring globally relevant information.

\subsection{CD Methods Based on Attention Mechanism}
In bi-temporal RS images, changes in surface objects often occur in localized regions, while other areas may exhibit color variations due to weather or seasonal effects, which can be regarded as noise outside of critical detections. The characteristic of attention mechanisms is to focus the model's attention on change-sensitive areas, effectively reducing the impact of noise regions. In recent years, numerous attention-based CD methods have emerged.

The adaptive spectral and spatial attention mechanism (S$^2$AN) \cite{33} integrates a spectral attention module that directly computes attention for each input channel with a gaussian spatial attention module that samples from an adaptive gaussian distribution, effectively suppressing spectral and spatial information unrelated to CD, thereby achieving noise reduction and improving model accuracy and generalization capability. AGCDetNet \cite{34} incorporates both spatial and channel attention into the model to enhance the discernibility of objects and backgrounds in change regions, resulting in commendable performance. The STADE-CDNet \cite{35} introduces a CD difference enhancement module, enhancing the network model with differential feature attributes through training layers to mitigate the issue of class imbalance.

The dual attention guided multi-scale feature fusion mechanism \cite{36} extracts features at different levels and combines spatial and channel attention with a three-branch fusion structure, effectively reducing background noise interference. Agent attention efficiently combines the high-performance softmax attention with low-computational linear attention, harnessing the benefits of both attention mechanisms. By reallocating feature map weights, the attention mechanism emphasizes key information, enhancing detection in change regions of bi-temporal images. Thus, an appropriate attention mechanism will assist the model in achieving better performance and results in CD tasks.

\begin{figure}[!t]
	\centering
	\includegraphics[width=3.5in]{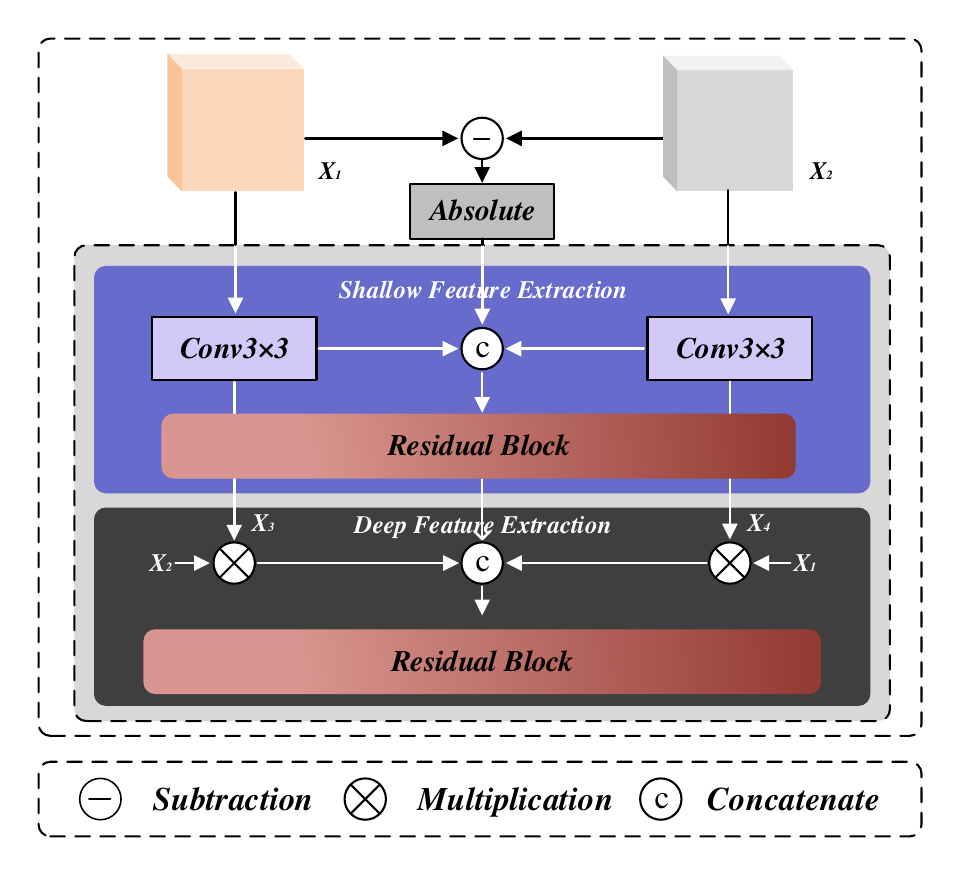}
	\caption{Shows the architecture of PEFM. It jointly processes the differential images and bi-temporal images, with shallow and deep features extracted based on residual structure. This progressive approach enhances the image fusion effect, thereby reducing the impact of pseudo-changes and shadow noise.}
	\label{fig3}
\end{figure}

\subsection{CD Methods Based on Transformer}
Transformers are primarily designed for sequence tasks, and their powerful capability to model long-range dependencies has garnered significant attention in computer vision tasks. The successful application of ViT has led to a proliferation of transformer variants across various visual domains \cite{37,38,39,40}. Leveraging the advantages of transformers in modeling long-distance dependencies can aid models in learning more global features of change regions. For instance, BIT \cite{41} inputs the features of cascaded bi-temporal images into a transformer for feature refinement.

CDViT \cite{42} further enhances the attention mechanism of transformers by initially employing spatial self-attention, which implicitly incorporates spatial information into each token, followed by spatiotemporal self-attention. This combination allows the learned tokens to encompass global contextual information. DiFormer \cite{43} introduces a difference evaluation module based on token exchanges, emphasizing the inconsistencies between change regions and surrounding contextual information, thereby highlighting the differences between bi-temporal images. EATDer \cite{44} addresses the significant impact of edge detection on mapping quality in change regions, designing an adaptive ViT module for siamese encoders and an integrated edge aware decoder that optimizes edge detection results based on refined features.

Various transformer methods based on siamese structures have demonstrated significant advantages in identifying subtle changes. By implementing weight sharing, these methods achieve symmetry in processing, mitigating additional biases during feature extraction while reducing the impact of pseudo-changes and focusing on authentic change regions, thus more effectively capturing the differences between bi-temporal images.

\begin{figure}[!t]
	\centering
	\includegraphics[width=3.5in]{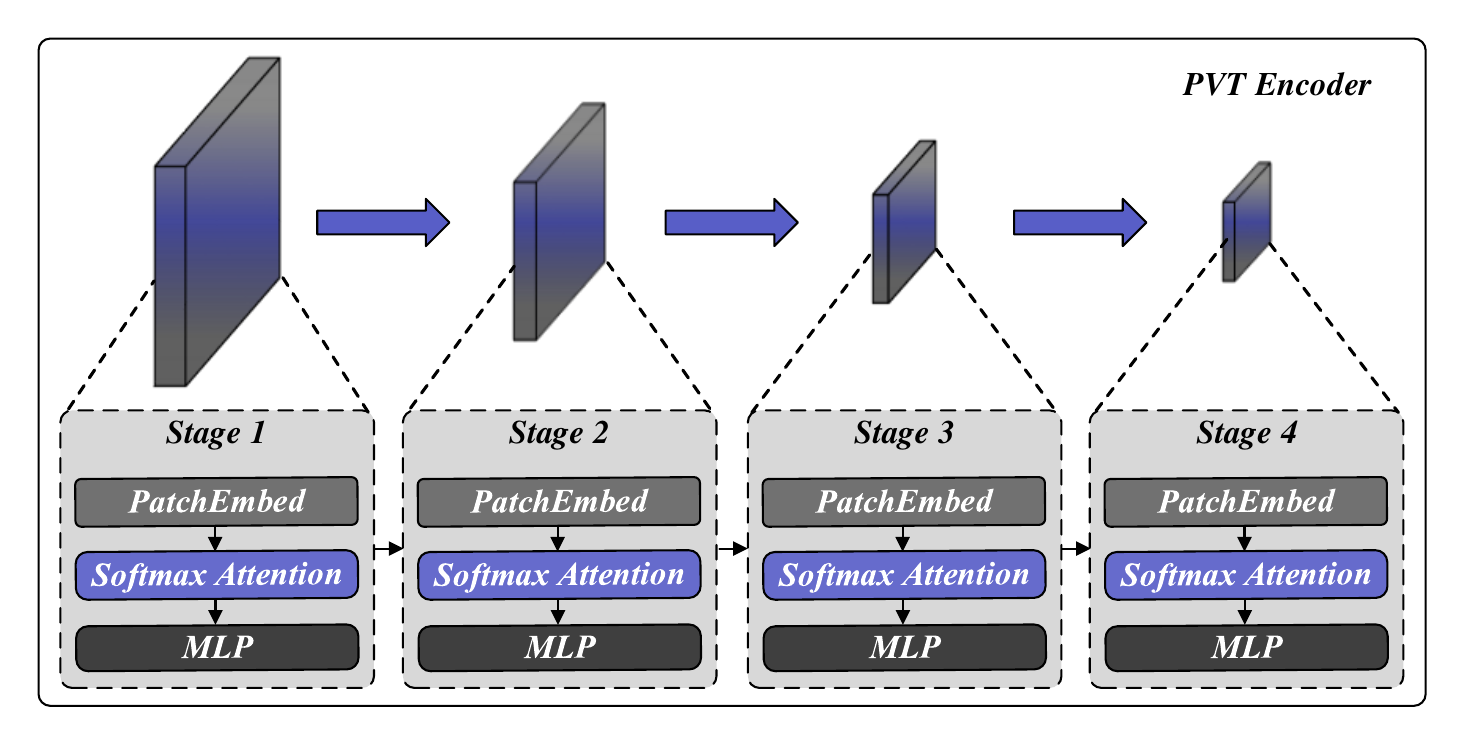}
	\caption{The overall architecture diagram of the PVT encoder.}
	\label{figpvt}
\end{figure}

\section{METHODOLOGY}
We will first introduce the overall architecture of our model, followed by a discussion of the importance of the PEFM in the network fusion component. Subsequently, we will provide an overview of the DCFM and its critical role in localizing change regions and mitigating noise.

\subsection{Overview Architecture}
\begin{figure}[!t]
	\centering
	\includegraphics[width=3.5in]{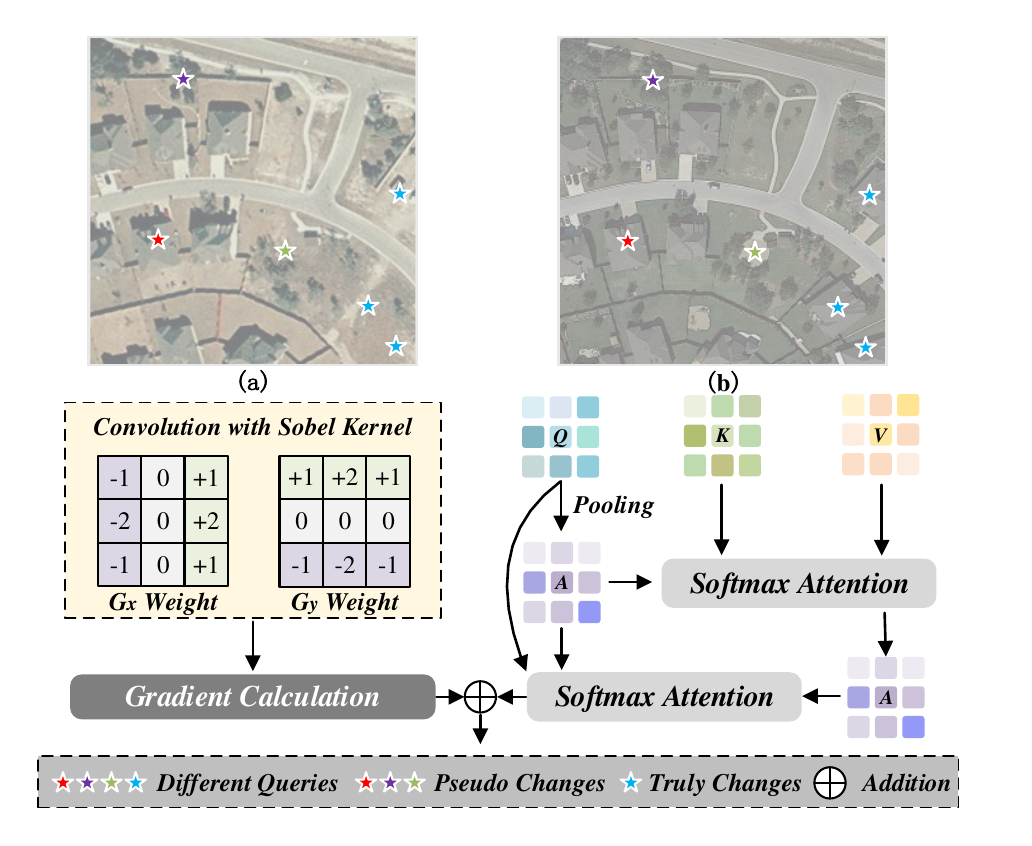}
	\caption{Shows the architecture of DCFM. In images (a) and (b), the red, purple, and green stars represent pseudo-change areas in the bi-temporal images, while the blue stars indicate real change areas. The agent attention mechanism focuses on the change regions and distinguishes pseudo-changes, while also addressing the shadows produced by buildings. The incorporation of edge detection algorithms further suppresses shadow noise from the buildings.}
	\label{fig4}
\end{figure}

Fig. \ref{fig2} illustrates the overall structure of our DFPF-Net. We first employ a PVT to construct a weight-sharing siamese architecture as the encoder. Subsequently, we input multi-scale feature information into the PEFM, which utilizes a dual residual structure to combine shallow and deep feature extraction to obtain the differential features of the bi-temporal images. The details of the PEFM are shown in Fig. \ref{fig3}. Next, we introduce the DCFM, which focuses on dynamically capturing the change regions in the bi-temporal images by reallocating weights to enhance CD performance. The details of the DCFM are shown in Fig. \ref{fig4}. Finally, we employ an attention-guided multiscale decoder to achieve effective fusion of hierarchical differential features.

\subsection{PVT Encoder}
The structure diagram of PVT is shown in Fig. \ref{figpvt}. When an image passes through a layer of the PVT, the patch embedding process divides the image into equally shaped small patches and transforms them into a one-dimensional sequence of vector representations. These vectors are then processed by the attention mechanism and the multilayer perceptron (MLP) within the transformer to extract fine-grained feature information. To effectively identify the change regions in bi-temporal images, we construct a siamese network to achieve weight sharing. Specifically, PVT incorporates a pyramid structure that builds features at different scales through progressive spatial downsampling, enhancing the model's ability to understand feature information and thereby better represent the image characteristics.

\begin{equation}\label{eq:1}
	X_{ij}=E_{i}\left(X_{i}\right)
\end{equation}
Where $X_{i}$ are the input images, and $E_{i}$ are the PVT encoders, where $i$ denotes the two parallel networks processing the bi-temporal images. $X_{ij}$ are the feature maps processed by the encoders, and $j$ represents the four scales of the PVT.

\subsection{PEFM}
As shown in Fig. \ref{fig3}, the PEFM is built on a dual-layer residual structure that enhances feature fusion of bi-temporal images, enabling effective change detection. The residual structure serves as the foundational layer for this module, providing stability in gradient computation, which ensures model stability during the training of deep networks. By utilizing skip connections, early features are directly passed to subsequent structures, offering a diversity of features to the model. It is precisely because of the inherent advantage of the residual structure in stabilizing training within deep networks that we can construct a progressive fusion network framed by dual residual structures. Specifically, we preprocess the bi-temporal images processed by PVT encoder to obtain $X_{ij}$, concatenate them with the difference image, and then pass them through the first residual layer $R_{1}$ to achieve initial feature enhancement, resulting in a fused feature map $X_{Shallow}$ containing shallow features.

\begin{equation}\label{eq:5}
	X^{'}_{ij}=Conv\left(X_{ij}\right)
\end{equation}

\begin{equation}\label{eq:7}
	X_{Shallow}=R_{1}\left(Cat\left(X_{1j},X_{2j},\left | X_{2j} - X_{1j}\right |\right)\right)
\end{equation}

Subsequently, we employ the concept of cross-attention to multiply the initially denoised bi-temporal images $X_{1j}^{'}$ with $X_{2j}$ and $X_{2j}^{'}$ with $X_{1j}$ resulting in $X_{Cross_{1}}$ and $X_{Cross_{2}}$. This process facilitates cross-temporal and spatial interaction of feature information, endowing $X_{1j}$ with change perception capabilities and $X_{2j}$ with difference detection capabilities. Following this, we concatenate these results with the fused image processed through the residual structure and pass them through the second residual layer $R_{2}$ for secondary feature enhancement, yielding the fused feature map $X_{Deep}$, which contains deep features. The PEFM progressively extracts hierarchical features from shallow to deep, further enhancing the understanding of overall scene changes and relationships among targets, thereby aiding the model in more accurately detecting large-scale change areas.

\begin{equation}\label{eq:8}
	X_{Cross_{1}}=X_{1j}^{'}\times X_{2j}
\end{equation}

\begin{equation}\label{eq:9}
	X_{Cross_{2}}=X_{2j}^{'}\times X_{1j}
\end{equation}

\begin{equation}\label{eq:10}
	X_{Deep}=R_{2}\left(Cat\left(X_{Cross_{1}} ,X_{Cross_{2}}, X_{Shallow}\right)\right)
\end{equation}

\subsection{DCFM}
As shown in Fig. \ref{fig4}, the DCFM integrates agent attention and edge detection algorithms to dynamically focus on change regions. The agent attention mechanism leverages softmax attention to capture long-range dependencies in the image while incorporating linear attention to reduce computational costs. By reallocating weights, this enhances the model’s responsiveness to significant change areas. In detail, agent attention first introduces $A$ as a proxy for $Q$, participating in the initial softmax attention process in a linear manner. Then, both the original $A$ and the aggregated $A$ act as proxies for $K$ and $V$ in the second softmax attention, achieving ssoftmax attention with linear time complexity. This dual-attention approach enables DCFM to refine focus on change-sensitive areas efficiently, supporting precise differentiation between true and pseudo-change areas.
\begin{equation}\label{eq:11}
	O_{i}^{S}=\sum_{j=1}^{N} \frac{\operatorname{S}\left(Q_{i}, K_{j}\right)}{\sum_{j=1}^{N} \operatorname{S}\left(Q_{i}, K_{j}\right)} V_{j}
\end{equation}

\begin{equation}\label{eq:12}
	O_{i}^{\phi}=\frac{\sum_{j=1}^{N}\left(\phi\left(Q_{i}\right) \phi\left(K_{j}\right)^{T}\right) V_{j}}{\sum_{j=1}^{N}\left(\phi\left(Q_{i}\right) \phi\left(K_{j}\right)^{T}\right)}
\end{equation}

\begin{equation}\label{eq:13}
	O_{i}^{A}=\sum_{j=1}^{N} \frac{\operatorname{S}\left(Q_{i}, K_{j},A\right)}{\sum_{j=1}^{N} \operatorname{S}\left(Q_{i}, K_{j},A\right)} V_{j}
\end{equation}
where $Q, K, V \in R^{N \times C}$ denote query, key and value matrices. $S\left(\cdot \right)$ measures the similarity between queries. Where $A \in R^{N \times C}$ are agent tokens containing the application of softmax operations on linear attention. $O_{S}$ denotes softmax attention, $O_{\phi}$ denotes linear attention and $O_{A}$ denotes agent attention.

The edge detection algorithm uses the sobel operator to describe the gradient changes in the horizontal and vertical directions of the image, thereby detecting edge information. Let the horizontal gradient after differentiation be $G_{x}$ and the vertical gradient be $G_{y}$, the approximate gradient $G$ can be mathematically expressed as follows:
\begin{equation}\label{eq:14}
	G=\sqrt{G_{x}^{2}+G_{y}^{2}}
\end{equation}

In Fig. \ref{fig4}, different colored stars represent the queries of the agent attention mechanism for various types of change regions, each with distinct representativeness. The purple star indicates changes in trees, the red star represents pseudo-changes in buildings, and the green star signifies other surface changes. All three types represent pseudo-change regions, which are global noise disturbances. The blue star denotes true change regions on a global scale. Factors like season and lighting can cause varying degrees of shadow around buildings, affecting the edge localization of targets, which is a local noise disturbance. To address these two types of noise interference, we combine the agent attention mechanism with a locally focused enhanced edge detection algorithm, enhancing the model's ability to locate differential targets and accurately segment the local edges of target areas.

\subsection{Cross Scale Interaction Decoder}
To correspond with the pyramid structure of the PVT encoder, we construct an attention-guided decoder using a cross-scale feature interaction structure. This involves upsampling low-dimensional features and aligning them with high-dimensional features. We then combine these features using a convolution-based attention mechanism to weight and fuse the different dimensions. Finally, we employ multiple convolutional blocks and residual connections to enhance the model's stability.

\begin{algorithm}[t]
	\caption{The implementation process of DFPF-Net}\label{alg:alg1}
	\begin{algorithmic}
		\STATE
		\STATE {\textbf{Input:}}\ $In_1,In_2\ (bitemporal\ image)$
		\STATE {\textbf{Output:}}\ $O_f$
		\STATE // $step1:Extract\ hierarchical\ features$
		\FOR{$i\ in\ \{1,2\}$}
		\FOR{$j\ in\ \{1,2,3,4\}$}
		\STATE \hspace{0.5cm}$\mathbf{F_{i,j}} = Siamese\ PVT \ Encoder(In_i)$
		\ENDFOR
		\ENDFOR
		\STATE // $step2:Progressive\ enhanced\ features\ fusion$
		\FOR{$j\ in\ \{1,2,3,4\}$}
		\STATE \hspace{0.5cm}$\mathbf{{F}^{S}_{j}} = Shallow(\mathbf{F_{1,j}}, \mathbf{F_{2,j}}, | \mathbf{F_{2,j}}-\mathbf{F_{1,j} |})$
		\STATE \hspace{0.5cm}$\mathbf{{F}^{D}_{j}} = Deep(\mathbf{F_{1,j}}, \mathbf{F_{2,j}}, \mathbf{{F}^{S}_{j}})$
		\ENDFOR
		\STATE // $step3:Dynamic\ change\ focus$
		\FOR{$j\ in\ \{1,2,3,4\}$}
		\STATE \hspace{0.5cm}$\mathbf{{F}^{A}_{j}} = AgentAttention(\mathbf{F}^{D}_{j})$
		\STATE \hspace{0.5cm}$\mathbf{{F}^{E}_{j}} = EdgeDetection(\mathbf{F}^{D}_{j})$
		\STATE \hspace{0.5cm}$\mathbf{{F}^{F}_{j}} = Fusion(\mathbf{F}^{A}_{j}, \mathbf{F}^{E}_{j})$
		\ENDFOR
		\STATE // $step4:Cross\ scale\ interaction$
		\FOR{$j\ in\ \{1,2,3\}$}
		\STATE $\mathbf{F^{up}} = Decoder(\mathbf{{F}^{F}_{j}}, \mathbf{{F}^{F}_{j+1}})$
		\STATE $\mathbf{{F}^{F}_{j+1}} =\mathbf{F^{up}}$
		\ENDFOR
		\STATE // $step5:Obtain\ the\ result$
		\STATE $O_{f} = Classify(\mathbf{F_{4}^{F}})$
	\end{algorithmic}
\end{algorithm}

\section{EXPERIMENTS}
We validate the effectiveness of DFPF-Net on four commonly used bi-temporal RS datasets: LEVIR-CD \cite{45}, WHU-CD \cite{46}, GZ-CD \cite{47}, and CDD \cite{57}.
\subsection{Datasets}
\subsubsection{LEVIR-CD Dataset}
LEVIR-CD is a widely used dataset for large-scale RS building change detection and has become an authoritative benchmark for evaluating CD algorithms. Each pair of images has a significant temporal gap, showcasing notable changes in surface objects and large spatiotemporal differences, making it effective for assessing the localization and segmentation capabilities of CD methods. The dataset includes a variety of scenes, such as rising and falling buildings, trees and roads with different color depths, and various building types like villas, high-rise apartments, and warehouses. It consists of 637 image pairs, with a resolution of 1024$ \times $1024. We randomly cropped the images to obtain independent 256×256 resolution images, which were then allocated to the training, testing, and validation sets in a 7:2:1 ratio.

\subsubsection{WHU-CD Dataset}
WHU-CD is a high-quality dataset used for building change detection, containing high-resolution images that include aerial photographs as well as large-scale satellite images with raster labels and vector maps. This dataset focuses on identifying and locating various types of buildings from RS images, featuring urban development changes and surface changes in areas affected by natural disasters from around the world. Accurately extracting building boundaries from complex and diverse backgrounds places high demands on the model's generalization ability. We cut the images into 256$ \times $256 pixel blocks, randomly allocating them into 6096 for training, 762 for validation, and 762 for testing.

\subsubsection{GZ-CD Dataset}
GZ-CD dataset is primarily used for CD tasks in RS images. It covers various complex scenes, including large-scale building changes and subtle changes that are difficult to detect, with challenges such as building shadows and overlapping areas, which pose significant difficulties for CD tasks. The dataset features buildings such as warehouses and densely packed strip buildings, aiming to evaluate the model's ability to handle details and accurately detect changes. We cropped the images to a size of 256$ \times $256 pixels, allocating 2834 images to the training set, 400 to the validation set, and 325 to the testing set.

\subsubsection{CDD Dataset}
CDD dataset is primarily used for building change detection severely affected by seasonal factors. Prominent seasonal changes include dense vegetation cover and post-snowfall scenes, where large-scale variations significantly increase background noise, making it challenging to effectively locate true change regions. These issues pose substantial challenges for model training. We cropped the images to a size of 256$ \times $256 pixels, including 10000 training images, 3000 validation images, and 3000 test images.

\begin{figure*}[t]
	\centering
	\includegraphics[width=7in]{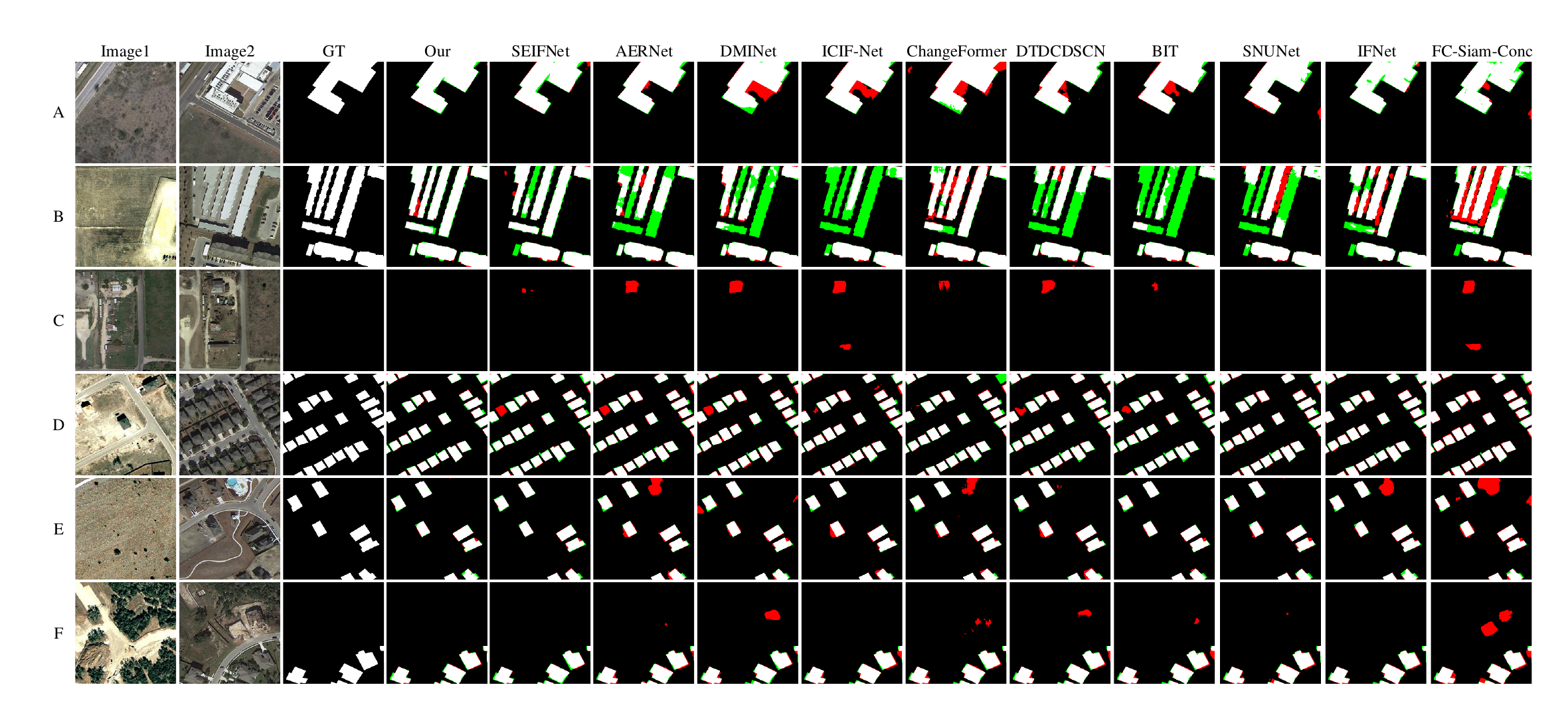}
	\caption{The comparison results of DFPF-Net with mainstream methods on the LEVIR-CD dataset are presented. In this figure, the green areas indicate parts of the prediction that are missing compared to the GT, while the red areas represent parts of the prediction that are redundant compared to the GT.}
	\label{fig5}
\end{figure*}

\subsection{Implementation Details}
All experiments were conducted on an Ubuntu 18.04 system with an NVIDIA TITAN RTX 24G GPU for model training in the PyTorch framework. We trained the model for 500 epochs, saving the best model after each epoch. The AdamW optimizer was used with an initial learning rate of $5e^{-4}$, and a cosine annealing schedule was applied for dynamic learning rate adjustment. Binary cross-entropy (BCE) loss was chosen for model optimization, defined as:
\begin{equation}\label{eq:15}
	\begin{aligned}
		\operatorname{BCELoss}\left(y_{\text {i}}, \hat{y}_{\text {i}}\right)&=-\frac{1}{N} \sum_{i=1}^{N}\left[y_{\text {i}} \log \left(\hat{y}_{\text {i}}\right)	\right.	\\
		&\left.+\left(1-y_{\text {i}}\right) \log \left(1-\hat{y}_{\text {i}}\right)\right]
	\end{aligned}
\end{equation}
where N denotes the number of pixels, $y_{i}$ represents the true value, and $\hat{y_{i}}$ denotes the predicted value.

\subsection{Evaluation Metrics}
We evaluate our model against other CD methods using four standard metrics: F1, IoU, precision, and recall. The F1 balances precision and recall, offering a comprehensive measure of performance, particularly in imbalanced classes. IoU quantifies the overlap between the predicted and ground truth masks, assessing the accuracy of change detection. Precision and recall focus on the model's effectiveness in identifying positive samples. The specific mathematical expressions for these metrics are as follows:
\begin{equation}\label{eq:16}
F1=\frac{2TP}{2TP+FP+FN}
\end{equation}

\begin{equation}\label{eq:17}
IoU=\frac{TP}{FP+FN+TP}
\end{equation}

\begin{equation}\label{eq:18}
Precision=\frac{TP}{TP+FP}
\end{equation}

\begin{equation}\label{eq:19}
Recall=\frac{TP}{TP+FN}
\end{equation}
Where TP represents true positives, FP represents false positives, FN represents false negatives, and TN represents true negatives.

\begin{figure*}[h]
	\centering
	\includegraphics[width=7in]{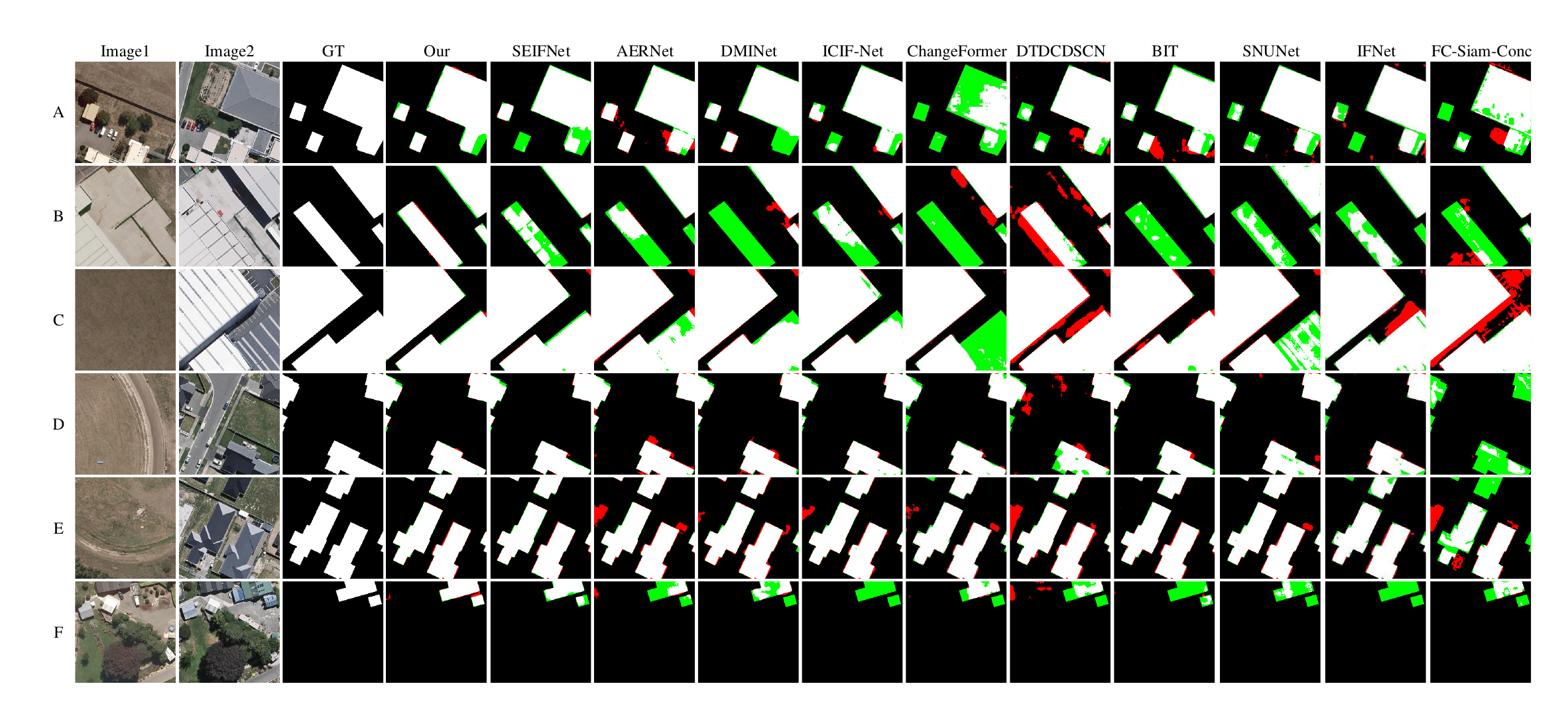}
	\caption{The comparison results of DFPF-Net with mainstream methods on the WHU-CD dataset are presented.}
	\label{fig6}
\end{figure*}

\begin{table}[h]
	\caption{Indicator results for the LEVIR-CD dataset. Red color represents the best results and blue color represents the second best results (\%).\label{tab1}}
	\centering
	\renewcommand\arraystretch{1.2}
	\resizebox{0.49\textwidth}{!}
	{
		\begin{tabular}{ccccc}
			\hline
			Method        & F1 & IoU & Precision & Recall \\
			\hline
			FC-Siam-Conc   & 81.77 & 69.16 & 84.17 & 79.49 \\
			IFNet          & 88.13 & 78.77 & 94.02 & 82.93 \\
			SNUNet         & 88.16 & 78.83 & 89.18 & 87.17 \\
			BIT            & 89.31 & 80.68 & 89.24 & 89.37 \\
			DTCDSCN        & 87.67 & 78.05 & 88.53 & 86.83 \\
			ChangeFormer   & 90.40 & 82.48 & 92.05 & 88.80 \\
			ICIF-Net       & \textcolor{blue}{91.18} & \textcolor{blue}{83.85} & 91.13 & \textcolor{blue}{90.57} \\
			DMINet         & 90.71 & 82.99 & 92.52 & 89.95 \\
			AERNet         & 90.78 & 83.11 & 89.97 & \textcolor{red}{91.59} \\
			SEIFNet        & 90.86 & 83.25 & \textcolor{blue}{94.29} & 87.67 \\
			Ours           & \textcolor{red}{91.77} & \textcolor{red}{84.80} & \textcolor{red}{94.33} & 89.35 \\
			\hline
		\end{tabular}
	}
\end{table}

\begin{table}[h]
	\caption{Indicator results for the WHU-CD dataset. Red color represents the best results and blue color represents the second best results (\%).\label{tab2}}
	\centering
	\renewcommand\arraystretch{1.2}
	\resizebox{0.49\textwidth}{!}
	{
		\begin{tabular}{ccccc}
			\hline
			Method          & F1 & IoU & Precision & Recall \\
			\hline
			FC-Siam-Conc   & 72.61 & 56.99 & 75.89 & 69.30 \\
			IFNet          & 83.40 & 71.52 & \textcolor{red}{96.91} & 73.19 \\
			SNUNet         & 88.34 & 79.11 & 91.34 & 85.53 \\
			BIT            & 87.47 & 77.73 & 88.71 & 86.27 \\
			DTCDSCN        & 90.48 & 82.62 & 91.84 & 89.16 \\
			ChangeFormer   & 86.88 & 76.81 & 88.50 & 85.33 \\
			ICIF-Net       & 90.77 & 83.09 & 92.93 & 88.70 \\
			DMINet         & 91.49 & 84.31 & 92.65 & 90.35 \\
			AERNet         & 92.18 & 85.49 & 92.47 & \textcolor{blue}{91.89} \\
			SEIFNet         & \textcolor{blue}{93.29} & \textcolor{blue}{87.43} & 93.99 & \textcolor{red}{92.61} \\
			Ours           & \textcolor{red}{93.79} & \textcolor{red}{88.30} & \textcolor{blue}{96.04} & 91.64 \\
			\hline
		\end{tabular}
	}
\end{table}

\subsection{Comparative Experiments}
We selected a total of nine mainstream CD methods for comparison with our approach. FC-Siam-Conc \cite{27} proposed three fully convolutional neural network architectures to refine feature information through stacked feature blocks. IFNet \cite{29} feeds the deep features extracted by the model into a depth-supervised differential discrimination network, combining multi-scale deep features and image difference features via an attention module to reconstruct the change map. SNUNet \cite{30} introduces a densely connected siamese network, focusing on fine-grained shallow features and employing a channel attention module to refine features across various semantic levels. DTCDSCN \cite{48} presented a deep siamese convolutional network with dual task constraints, which helps learn discriminative object-level features while introducing dual attention modules to enhance feature representation by utilizing the dependencies between channel and spatial locations. AER-Net \cite{49} presents an attention-guided edge refinement network that aggregates global context features across multiple layers, with an edge optimization module to improve sensitivity to edge change areas. DMINet \cite{50} introduces a dual-branch multi-level network that utilizes self-attention and cross-attention to guide global feature distribution, improving inter-layer information coupling and reducing noise interference. BIT \cite{41} utilized a transformer encoder to model the context of bi-temporal RS images in vector form, feeding back contextually informed tokens to pixel space to optimize the feature information in the decoding section. ChangFormer \cite{51} proposed a hierarchical transformer structure combined with an MLP decoder to construct a siamese network, thereby establishing multi-scale long-range dependencies. ICIF-Net \cite{52} leverages the strengths of CNN and transformer by employing a linear convolutional attention module to enable feature interaction across both architectures, effectively harnessing their combined potential. SEIFNet \cite{56} designs a spatiotemporal difference enhancement module and an adaptive context fusion module, further exploring temporal differences and multi-scale features to enhance the feature representation ability for changing objects, while integrating inter-layer features to complete decoding.

\begin{figure*}[h]
	\centering
	\includegraphics[width=7.2in]{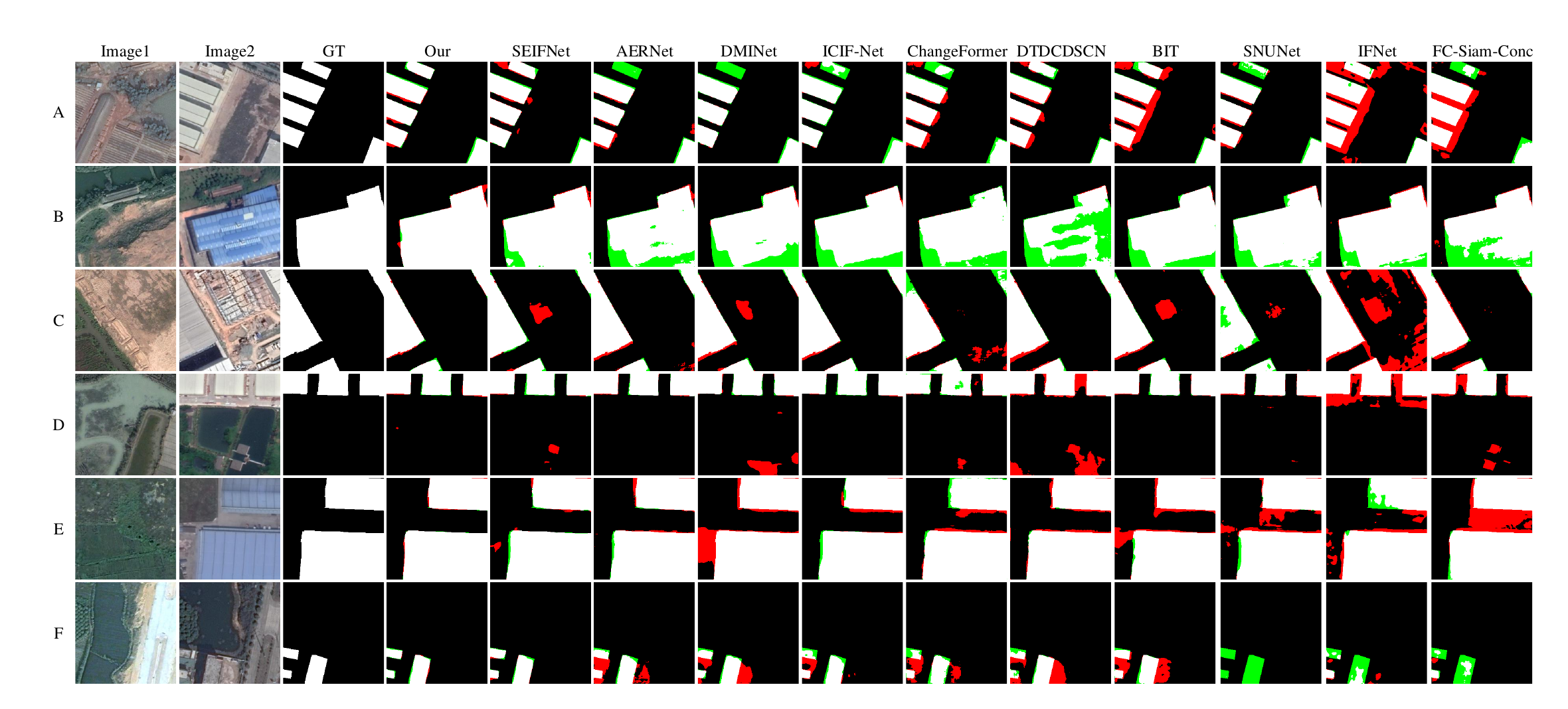}
	\caption{The comparison results of DFPF-Net with mainstream methods on the GZ-CD dataset are presented.}
	\label{fig7}
\end{figure*}

As shown in Fig. \ref{fig5}, we compared the prediction maps generated by DFPF-Net with those of other methods, selecting representative samples A-F from the LEVIR-CD dataset. Sample A demonstrates the prediction results for a large area building, where DMINet and ICIF-Net are affected by building shadows, leading to redundancies with unusual shapes. Sample B displays the CD results for a cluster of strip-shaped buildings, where various mainstream methods show significant missing predictions. Sample C contains no real change areas but still exhibits pseudo-changes. Sample D features densely packed small buildings with interference from pseudo-changes in the building category. In samples E and F, various mainstream methods produce different degrees of redundant predictions due to significant background color differences. The advantages of our proposed method are prominently highlighted in these samples. We achieved stable predictions for the main areas of change using the pyramid vision transformer, resulting in only minimal missing predictions. Our PEFM supports the transformer in achieving structured multi-level fusion, effectively addressing the pseudo-change noise in small target areas. At the same time, we successfully countered the noise generated by building shadows by organically combining dynamic attention focusing with edge detection algorithms. Table. \ref{tab1} visually presents the metrics for various methods on the LEVIR-CD dataset, showing that our DFPF-Net outperforms the second-place ICIF-Net by 0.59\% and 0.95\% in F1 and IoU, respectively.

\begin{table}[h]
	\caption{Indicator results for the GZ-CD dataset. Red color represents the best results and blue color represents the second best results (\%).\label{tab3}}
	\centering
	\renewcommand\arraystretch{1.2}
	\resizebox{0.49\textwidth}{!}
	{
		\begin{tabular}{ccccc}
			\hline
			Method          & F1 & IoU & Precision & Recall \\
			\hline
			FC-Siam-Conc   & 74.23 & 59.03 & 80.37 & 68.97 \\
			IFNet          & 82.15 & 69.71 & \textcolor{blue}{92.19} & 74.08 \\
			SNUNet         & 84.25 & 72.79 & 84.25 & 81.82 \\
			BIT            & 80.23 & 66.99 & 82.40 & 78.18 \\
			DTCDSCN        & 83.00 & 70.93 & 88.19 & 78.38 \\
			ChangeFormer   & 73.66 & 58.30 & 84.59 & 65.23 \\
			ICIF-Net       & 85.09 & 74.05 & 89.90 & 80.76 \\
			DMINet         & 81.98 & 69.46 & 87.92 & 76.79 \\
			AERNet         & 84.42 & 73.03 & 88.06 & 81.07 \\
			SEIFNet        & \textcolor{blue}{87.48} & \textcolor{blue}{77.75} & 89.64 & \textcolor{red}{85.43} \\
			Ours           & \textcolor{red}{87.83} & \textcolor{red}{78.30} & \textcolor{red}{93.06} & \textcolor{blue}{83.15} \\
			\hline
		\end{tabular}
	}
\end{table}

In Fig. \ref{fig6}, we compare the model prediction maps generated by DFPF-Net with the prediction results from other methods, selecting representative samples A-F from the WHU-CD dataset. Samples A, B, and C represent large-area building CD images, featuring relatively simple background information. However, in Sample B, the nearly identical background color results in DMINet and ChangeFormer completely missing the prediction for the building in the lower left corner. Samples D, E, and F involve building change detection with complex edge information, where the target areas are surrounded by strong guiding block and strip information. This strong interference noise severely affects AERNet and DTCDSCN. Our method remains unaffected by these issues, effectively handling noise information from similar background colors while ignoring strong guiding interference around the changing targets, and ensuring accurate predictions of complex edges. This demonstrates that our model possesses strong robustness. Table. \ref{tab2} provides a clear presentation of the metrics for various methods on the WHU-CD dataset. Our DFPF-Net outperforms the second-place SEIFNet by a significant margin of 0.5\% and 0.87\% in F1 and IoU, respectively.

\begin{figure*}[h]
	\centering
	\includegraphics[width=7in]{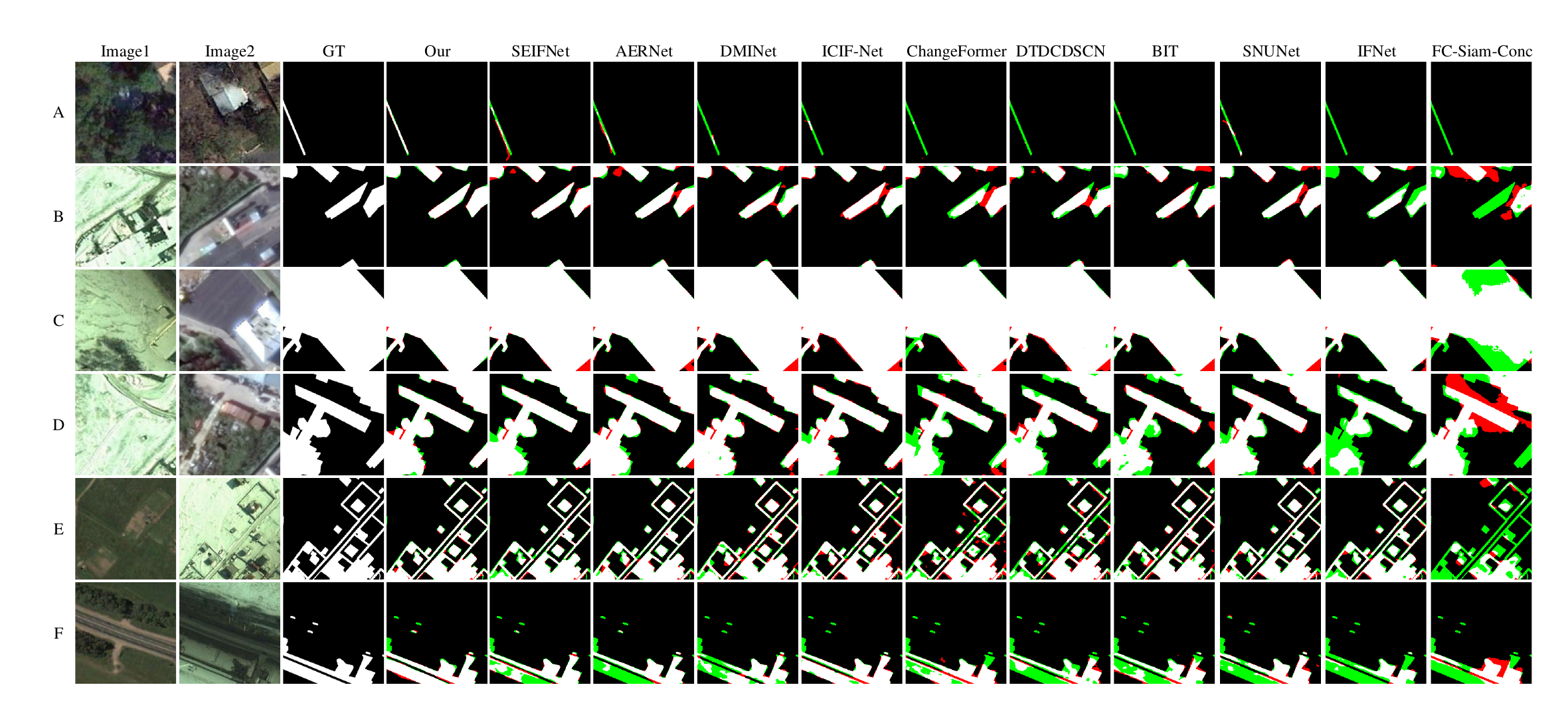}
	\caption{The comparison results of DFPF-Net with mainstream methods on the CDD dataset are presented.}
	\label{figCDD}
\end{figure*}

\begin{table}[h]
	\caption{Indicator results for the CDD dataset. Red color represents the best results and blue color represents the second best results (\%).\label{tabCDD}}
	\centering
	\renewcommand\arraystretch{1.2}
	\resizebox{0.49\textwidth}{!}
	{
		\begin{tabular}{ccccc}
			\hline
			Method          & F1 & IoU & Precision & Recall \\
			\hline
			FC-Siam-Conc   & 75.40 & 60.51 & 93.58 & 63.13 \\
			IFNet          & 91.65 & 84.59 & 99.07 & 85.27 \\
			SNUNet         & 93.62 & 88.00 & 98.67 & 89.06 \\
			BIT            & 92.59 & 86.19 & 97.52 & 88.12 \\
			DTCDSCN        & 91.52 & 84.36 & 96.71 & 86.85 \\
			ChangeFormer   & 90.73 & 83.03 & 97.35 & 84.95 \\
			ICIF-Net       & 93.73 & 88.20 & 97.21 & \textcolor{red}{90.50} \\
			DMINet         & 92.93 & 86.79 & 97.30 & 88.94 \\
			AERNet         & \textcolor{blue}{93.93} &\textcolor{blue}{88.56} & 99.05 & 89.31 \\
			SEIFNet        & 93.86 & 88.43 & \textcolor{red}{99.18} & 89.08 \\
			Ours           & \textcolor{red}{94.47} & \textcolor{red}{89.52} & \textcolor{blue}{99.13} & \textcolor{blue}{90.23} \\
			\hline
		\end{tabular}
	}
\end{table}

In Fig. \ref{fig7}, we compare the model prediction maps generated by DFPF-Net with the prediction results from other methods, selecting representative samples A-F from the GZ-CD dataset. Samples A, D, and F illustrate changes in small block-like buildings, where various methods exhibit irregular missing and redundant predictions to varying degrees. In some cases of redundant predictions, multiple buildings are even connected, indicating that the stability of the models in these methods still needs improvement. Sample B features buildings with multiple layers, where the entire structure that rises above the ground is the true change area, while the blue portion is only part of the true change area. This significant color difference and the potential layering of buildings result in missing predictions across all methods for this region. Our method consistently maintains accurate predictions of the main changing targets and good edge segmentation across different types of building detections. Even in addressing the unique changes presented in samples like B, it demonstrates strong feature extraction and analytical capabilities. Table. \ref{tab3} provides a clear presentation of the metrics for various methods on the GZ-CD dataset. Our DFPF-Net achieves impressive results, outperforming the second-place SEIFNet by 0.35\% and 0.55\% in F1 and IoU, respectively.

In Fig. \ref{figCDD}, we compared DFPF-Net's prediction maps with other methods using representative samples A-F from the CDD dataset. In Sample A, the detection of slender strip-shaped buildings in images contrasting lush vegetation and withered vegetation is challenging to discern with the naked eye, and other methods fail to accurately detect the change area. Our model overcomes this difficulty through a noise filtering mechanism. Samples B-E showcase comparison images with and without snow cover, highlighting irregular shapes, large areas, and intricate detailed change regions as representative images to demonstrate the superiority of our model. Table. \ref{tabCDD} provides a clear presentation of the metrics for various methods on the CDD dataset. Our DFPF-Net achieves impressive results, outperforming the second-place AERNet by 0.54\% and 0.96\% in F1 and IoU, respectively.

\begin{figure*}[!t]
	\centering
	\includegraphics[width=6.5in]{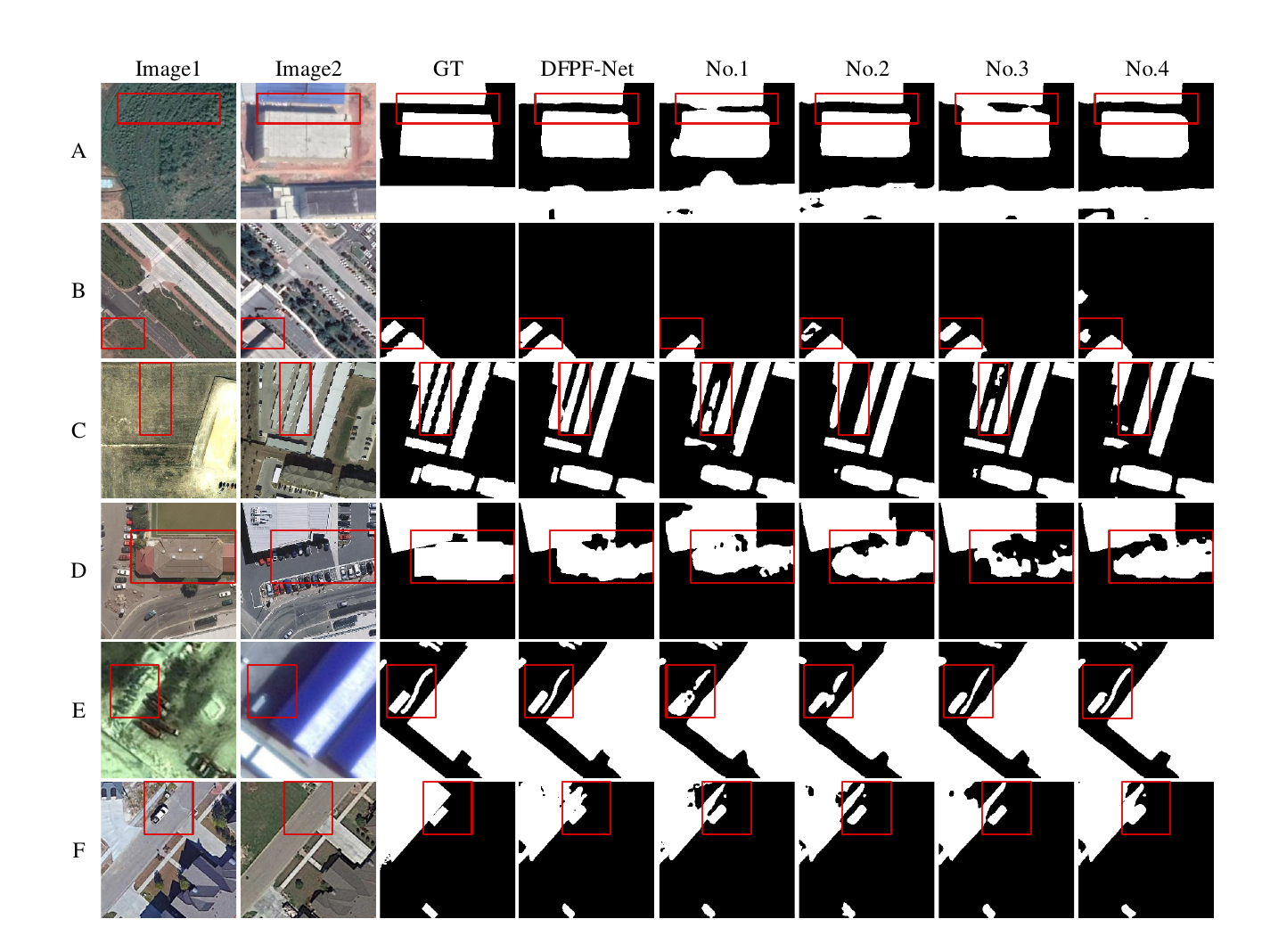}
	\caption{Ablation experiments of DFPF-Net on four datasets, with key focus areas in the samples highlighted using red boxes.}
	\label{fig8}
\end{figure*}

\begin{table*}[!ht]
	\centering
	\setlength{\tabcolsep}{9pt}
	\caption{\textrm{Overview of ablation results for DFPF-Net and metric outcomes on four datasets (\%).}}
	\begin{tabular}{ccccccccccccc}
		\hline
		\textbf{No.} & \textbf{PEFM} & \textbf{DCFM} & \textbf{DCFM-$\alpha$ } & \textbf{DCFM-$\beta$} & \multicolumn{2}{c}{\textbf{LEVIR-CD}} & \multicolumn{2}{c}{\textbf{WHU-CD}} & \multicolumn{2}{c}{\textbf{GZ-CD}} & \multicolumn{2}{c}{\textbf{CDD}} \\
		\cline{6-13}
		& & & & & \textbf{F1} & \textbf{IoU} & \textbf{F1} & \textbf{IoU} & \textbf{F1} & \textbf{IoU}  & \textbf{F1} & \textbf{IoU} \\
		\hline
		\textbf{1}   & $\times$ & \checkmark & \checkmark & \checkmark
		& 91.48 & 84.30 & 93.47 & 87.75 & 87.56 & 77.87 & 93.63 & 88.03 \\
		\textbf{2}   & \checkmark & $\times$ & \checkmark & \checkmark
		& 91.40 & 84.16 & 93.23 & 87.31 & 87.43 & 77.67 & 93.87 & 88.46 \\
		\textbf{3}   & $\times$ & \checkmark & $\times$ & \checkmark
		& 91.61 & 84.52 & 93.61 & 87.98 & 87.66 & 78.02 & 94.02 & 88.71 \\
		\textbf{4}   & $\times$ & \checkmark & \checkmark & $\times$
		& 91.65 & 84.59 & 93.59 & 87.96 & 87.60 & 77.93 & 93.94 & 88.58 \\
		\hline
	\end{tabular}
	\label{tab4}
\end{table*}

\begin{table}[h]
	\caption{Sensitivity analysis experiment of key model parameter settings on the LEVIR dataset (\%).\label{tablr}}
	\centering
	\renewcommand\arraystretch{1.2}
	\resizebox{0.49\textwidth}{!}
	{
		\begin{tabular}{ccccc}
			\hline
			Learning Rate        & F1 & IoU & Precision & Recall \\
			\hline
			1e-5       & 91.68 & 84.64 & 93.01 & 90.39 \\
			1e-4       & 91.73 & 84.72 & 92.87 & 90.61 \\
			5e-4       & 91.77 & 84.80 & 94.33 & 89.35 \\
			1e-3       & 91.64 & 84.57 & 93.03 & 90.29 \\
			5e-3       & 83.11 & 71.10 & 94.17 & 74.38 \\
			\hline
		\end{tabular}
	}
\end{table}

Overall, we selected prediction results from four datasets that reflect various scenarios in CD from different perspectives. We compared the predictions of DFPF-Net with those of mainstream methods, highlighting the effectiveness of our proposed transformer architecture, PEFM, and DCFM in removing noise related to pseudo-changes and building shadows. This comparison confirms that our model achieves advanced capabilities in differential discrimination.

\subsection{Ablation Experiments}
To validate the effectiveness of our DFPF-Net, we conducted ablation experiments by removing PEFM and DCFM. Additionally, to further demonstrate the impact of the attention mechanism and edge detection algorithm within DCFM, we set up two alternative versions: one without the attention mechanism, referred to as DCFM-$\alpha$, and the other without the edge detection algorithm, referred to as DCFM-$\beta$.

In Fig. \ref{fig8}, we selected representative samples from four datasets to demonstrate the roles and effectiveness of different components of DFPF-Net. In Sample A, building shadows exist between two change areas, and models No. 1 and No. 3 perform poorly, both lack the attention mechanism in DCFM, confirming that agent attention effectively identifies shadow regions within change target areas, thereby suppressing shadow noise. In Sample B, No. 1 fails to detect the target area, indicating that the absence of the attention mechanism leads to loss of the model's ability to capture global features, resulting in missing change detections. Sample No. 4 exhibits missing edges and redundant predictions, the edge detection algorithm effectively outlines the edge features of change areas, and combined with the global attention mechanism, refines the predicted edge results while ensuring overall target accuracy. In Sample C, both No. 2 and No. 4 fail to predict changes in strip-like buildings. No. 2 removed the progressive fusion mechanism, using concatenation to replace PEFM to maintain model integrity, but this change resulted in poor fusion of bi-temporal images and ineffective extraction of difference features, further compromising the enhancement of change focus. Although No. 4 includes the attention mechanism, it still fails to capture all differences globally. The extraction of boundary features complements the attention mechanism, promoting the capture of difference features and boundary information. In Samples D, E, and F, the predicted target buildings show internal tearing, leading to mispredictions as multiple target areas. Our complete model demonstrates strong robustness in handling such issues.

The prediction results in Fig. \ref{fig8} demonstrate our model's excellent ability to distinguish between pseudo-changes and suppress the influence of shadow noise. The data in Table. \ref{tab4} visually showcase the performance enhancements provided by the progressive fusion mechanism and dynamic change focusing mechanism on the baseline model. Overall, our ablation experiments conducted from multiple angles confirmed the effectiveness and necessity of each component, solving issues while effectively improving model accuracy and highlighting the superiority of DFPF-Net.

To intuitively demonstrate the processing of images by DFPF-Net and the effectiveness of our proposed methods, we conducted experimental analysis on the prediction results at the heatmap level. We set heatmap checkpoints at various stages of the model, including the encoder phase, Stage 1 and Stage 2 of the PEFM, and the DCFM stage, to output the feature map effects corresponding to each stage. The color differences within the regions clearly indicate the unique feature representations of the images at each stage.

\begin{table*}[!ht]
	\centering
	\setlength{\tabcolsep}{10pt}
	\caption{\textrm{Comparison of Parameters$\left( M\right)$, FLOPs$\left( G\right)$, and the basic time for the model in each epoch T$\left( s\right)$, along with a brief description of the comparison methods for F1(\%) and IoU(\%) performance.}}
	\begin{tabular}{cccccccccccc}
		\hline
		\textbf{Method} & \textbf{Params} & \textbf{FLOPs} & \textbf{T} & \multicolumn{2}{c}{\textbf{LEVIR-CD}} & \multicolumn{2}{c}{\textbf{WHU-CD}} & \multicolumn{2}{c}{\textbf{GZ-CD}} & \multicolumn{2}{c}{\textbf{CDD}} \\
		\cline{5-12}
		& & & & \textbf{F1} & \textbf{IoU} & \textbf{F1} & \textbf{IoU} & \textbf{F1} & \textbf{IoU} & \textbf{F1} & \textbf{IoU} \\
		\hline
		\textbf{FC-Siam-Conc}   & 1.55  & 5.32   & 0.25 & 81.77 & 69.16 & 72.61 & 56.99 & 74.23 & 59.03 & 75.40 & 60.51 \\
		\textbf{IFNet}          & 50.71 & 82.35  & 0.90 & 88.13 & 78.77 & 83.40 & 71.52 & 82.15 & 69.71 & 91.65 & 84.59 \\
		\textbf{SNUNet}         & 12.03 & 54.88  & 0.92 & 88.16 & 78.83 & 88.34 & 79.11 & 84.25 & 72.79 & 93.62 & 88.00 \\
		\textbf{BIT}            & 3.55  & 10.59  & 0.27 & 89.31 & 80.68 & 87.47 & 77.73 & 80.23 & 66.99 & 92.59 & 86.19 \\
		\textbf{DTCDSCN}        & 41.07 & 13.21  & 0.28 & 87.67 & 78.05 & 90.48 & 82.62 & 83.00 & 70.93 & 91.52 & 84.36 \\
		\textbf{ChangeFormer}   & 41.03 & 202.83 & 1.42 & 90.40 & 82.48 & 86.88 & 76.81 & 73.66 & 58.30 & 90.73 & 83.03 \\
		\textbf{ICIF-Net}       & 25.83 & 25.27  & 0.72 & 91.18 & 83.85 & 90.77 & 83.09 & 85.09 & 74.05 & 93.73 & 88.20 \\
		\textbf{DMINet}         & 6.24  & 14.55  & 1.19 & 90.71 & 82.99 & 91.49 & 84.31 & 81.98 & 69.46 & 92.93 & 86.79 \\
		\textbf{AERNet}         & 25.36 & 12.82  & 0.34 & 90.78 & 83.11 & 92.18 & 85.49 & 84.42 & 73.03 & 93.93 & 88.56 \\
		\textbf{SEIFNet}        & 8.37  & 27.9   & 0.22 & 90.86 & 83.25 & 93.68 & 88.11 & 87.48 & 77.75 & 93.86 & 88.43 \\
		\textbf{Ours}           & 46.67 & 16.89  & 0.64 & 91.77 & 84.80 & 93.79 & 88.30 & 87.83 & 78.30 & 94.47 & 89.52 \\
		\hline
	\end{tabular}
	\label{tab5}
\end{table*}

\begin{figure*}[!t]
	\centering
	\includegraphics[width=6.0in]{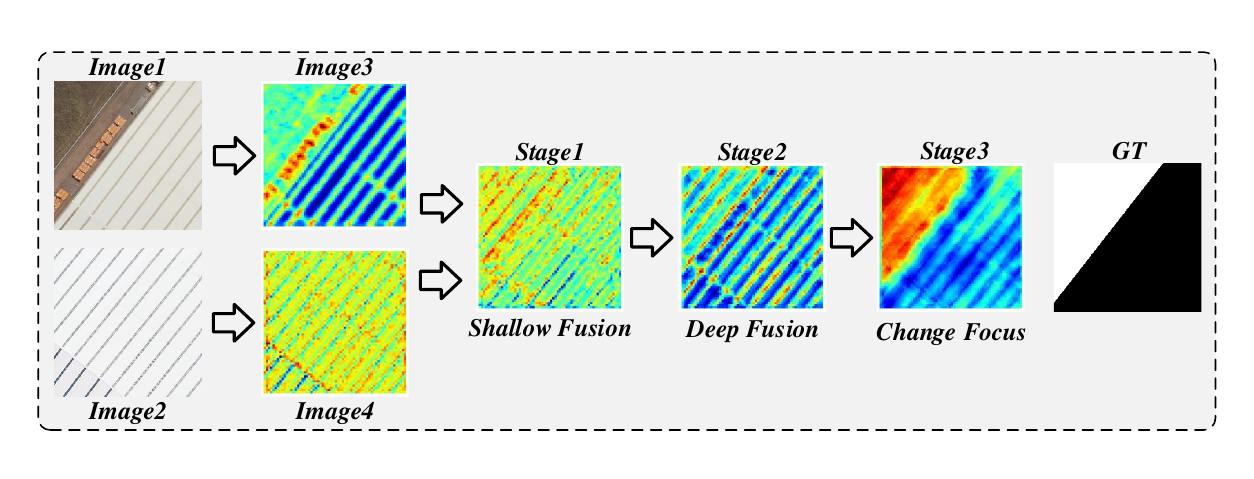}
	\caption{Display the heatmaps generated by DFPF-Net after key stages of processing.}
	\label{fig9}
\end{figure*}

As shown in Fig. \ref{fig9}, Image1 and Image2 represent the bi-temporal RS images, while Image3 and Image4 are the input images processed by the PVT encoder, each extracting initial features in different forms. Stage 1 indicates the shallow feature extraction of PEFM, where the red striped area reveals the initial fusion features of the image. Stage 2 represents the deep feature extraction of PEFM, the formation of deep blue areas contrasts with the red-green areas in the upper left, illustrating the progressive deep fusion features. Stage 3 signifies dynamic change aggregation, with the highlighted red areas indicating the effectiveness of agent attention in focusing on global difference regions. The bright edges adjacent to the red areas clearly separate the difference regions from the pseudo-change areas, emphasizing the necessity of the organic combination of the attention mechanism and edge detection algorithm in DFPF-Net.

To more comprehensively evaluate the robustness and generalization ability of the model, we added a sensitivity analysis experiment on key parameter settings in the ablation study section. This experiment conducts on the LEVIR dataset, recording the changes in various performance metrics during model training under different learning rates, as shown in Table. \ref{tablr}. By comparing the performance differences under different parameter settings and analyzing the experimental results, it observes that training the model with learning rates between 1e-5 and 1e-3 can approximate optimal performance. However, at smaller learning rates, the number of training epochs required to achieve the most effective performance increases significantly. Conversely, training with a larger learning rate of 5e-3, even after completing all epochs, resulted in performance far from optimal and encountered issues such as gradient explosion in later epochs. This highlights the importance of conducting sensitivity analysis on key parameters in the model.

\subsection{Parameters and FLOPs comparison}
As shown in Table. \ref{tab5}, we present a comparison between various methods and DFPF-Net in terms of parameters, FLOPs, and the time every epoch of model operation T, as well as a comparison of other methods in terms of F1 and IoU. The number of parameters reflects the model's memory requirements, while FLOPs estimate the computational workload during inference. Our model has 46.67M parameters, 16.89G FLOPs, and 0.64s for T. In comparison with other methods, DFPF-Net has a larger number of parameters, lower computational workload, and moderate runtime every epoch, while outperforming other methods in terms of F1 and IoU. This ensures that, with a moderate inference speed, our method maintains a superior position in terms of prediction performance and computational efficiency.

\begin{figure}[!t]
	\centering
	\includegraphics[width=3.5in]{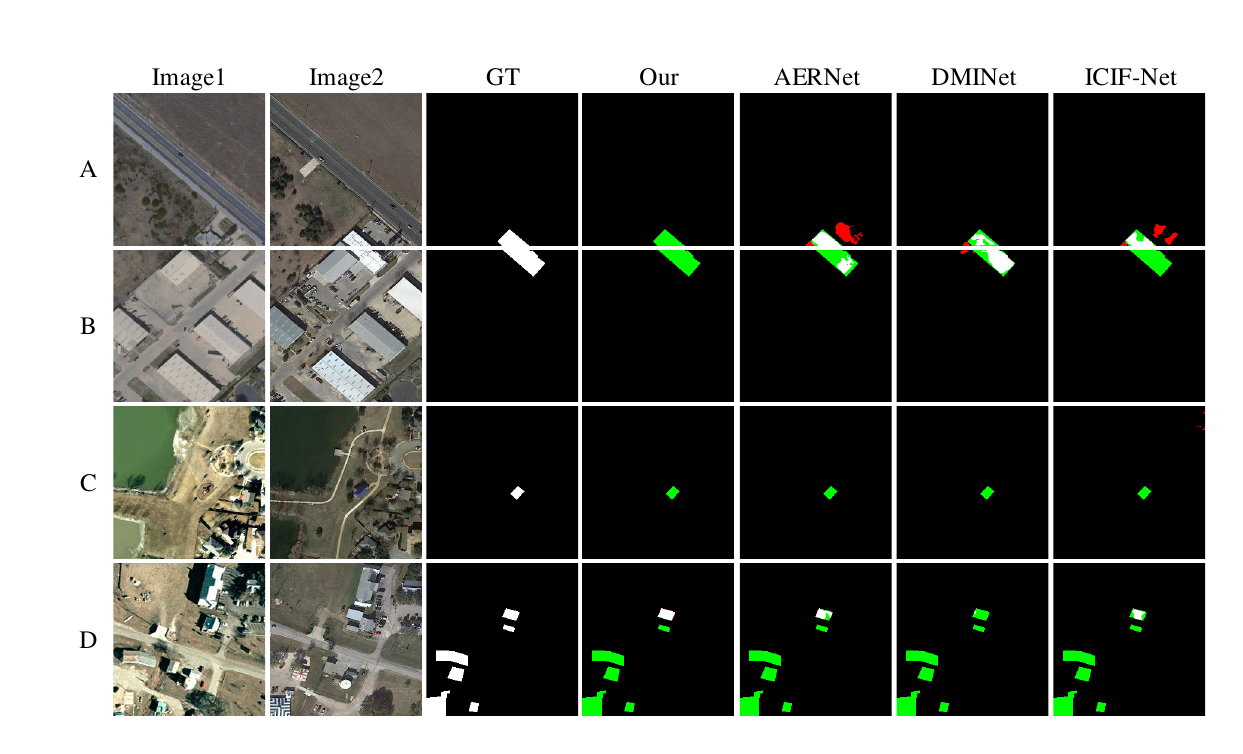}
	\caption{Display of missing detection for building changes with large color differences.}
	\label{fig10}
\end{figure}

\section{LIMITATIONS AND FUTURE WORK}
In extensive experiments across the four datasets, the prediction results obtained by DFPF-Net still exhibited varying degrees of redundancy and omissions compared to the ground truth. Our progressive fusion structure offers high flexibility and scalability, but the feature interaction process of bi-temporal images requires further exploration. We plan to enrich the fusion structure to achieve a better feature interaction process. The effective combination of the attention mechanism and edge detection algorithm has improved the model's accuracy, but their integration is relatively simple, leading to limitations in detecting differences in complex backgrounds. In addition, as shown in Fig. \ref{fig10}, we identified the category with the most severe missing predictions from a large number of experimental results. We found that changes in buildings with overly bright or dark colors were almost completely undetected, and this issue also exists in other mainstream methods. Therefore, addressing the missing detection of building changes caused by large color differences due to factors such as brightness and contrast is one of our future research directions.

\section{CONCLUSION}
In this paper, we employed a siamese network with a PVT encoder for CD in bi-temporal RS images to effectively distinguish between globally pervasive pseudo-change areas and genuine change regions. Specifically, the model extracts multi-scale feature information using PVT and leverages the PEFM, based on residual structures, for effective fusion of differential images and features at various levels. This fully utilizes the transformer's advantage in handling long-range dependencies, mitigating noise from globally prevalent pseudo-change areas while accurately locating genuine change regions. To further address the noise impact from global sources and local noise caused by building shadows due to lighting conditions, we designed the DCFM, which combines a high-performance attention mechanism with a low-computation-cost edge detection algorithm. This dynamic attention mechanism focuses on pinpointing difference areas, while the edge detection algorithm compensates for the transformer’s limitations in processing local edge information. Overall, we designed the PEFM to assist the PVT in feature fusion for initial localization of change areas, followed by the DCFM for further localization of genuine change regions while addressing local noise from building shadows. Finally, the decoder performs layer-wise interaction of feature information for upsampling. DFPF-Net effectively mitigates interference from pseudo-changes and building shadows, with experiments on four datasets demonstrating that our method achieves optimal results compared to  mainstream CD methods.

\section*{Acknowledgment}
This research was supported by the National Natural Science Foundation of China (61772319, 62272281, 62002200, 62202268), Shandong Natural Science Foundation of China (Grant no. ZR2023MF026, ZR2022MA076), Yantai science and technology innovation development plan(2023JCYJ040).


\bibliographystyle{IEEEtran}
\bibliography{references}{}

\end{document}